# XML Schema-Based Minification for Communication of Security Information and Event Management (SIEM) Systems in Cloud Environments

Bishoy Moussa
Information Technology Department
Faculty of Computers and
Information, Helwan University
Cairo, Egypt

Mahmoud Mostafa
Information Systems Department
Faculty of Computers and
Information, Helwan University
Cairo, Egypt

Mahmoud El-Khouly
Information Technology Department
Faculty of Computers and
Information, Helwan University
Cairo, Egypt

*Abstract*—XML-based communication governs most of today's systems communication, due to its capability of representing complex structural and hierarchical data. However, XML document structure is considered a huge and bulky data that can be reduced to minimize bandwidth usage, transmission time, and maximize performance. This contributes to a more efficient and utilized resource usage. In cloud environments, this affects the amount of money the consumer pays. Several techniques are used to achieve this goal. This paper discusses these techniques and proposes a new XML Schema-based Minification technique. The proposed technique works on XML Structure reduction using minification. The proposed technique provides a separation between the meaningful names and the underlying minified names, which enhances software/code readability. This technique is applied to Intrusion Detection Message Exchange Format (IDMEF) messages, as part of Security Information and Event Management (SIEM) system communication hosted on Microsoft Azure Cloud. Test results show message size reduction ranging from 8.15% to 50.34% in the raw message, without using time-consuming compression techniques. Adding GZip compression to the proposed technique produces 66.1% shorter message size compared to original XML messages.

*Keywords—XML; JSON; Minification; XML Schema; Cloud; Log; Communication; Compression; XMill; GZip; Code Generation; Code Readability*

## I. INTRODUCTION

XML-based communication governs most of today's systems communication, due to its capability of representing complex structural and hierarchical data. However, XML document structure is considered a huge and bulky data that can be reduced to minimize bandwidth usage, transmission time, and maximize performance. This contributes to a more efficient and utilized resource usage. In cloud environments, this affects the amount of money the consumer pays. Several techniques are used to achieve this goal. This paper discusses these techniques and proposes a new XML Schema-based Minification technique. The proposed technique works on XML Structure reduction using minification. The technique separates the original structure names from the minified names, to better achieve code readability while reducing data sent in the wire. This technique is applied to Intrusion Detection Message Exchange Format (IDMEF) messages, as part of Security Information and Event Management (SIEM) system communication hosted on Microsoft Azure Cloud.

This paper starts with an overview of the key concepts, required throughout the paper in section II. Section III presents related work. Then, section IV introduces the proposed system architecture and the minification process. After that, two experiments and test results are presented in section V. Conclusively, the proposed solution is discussed, and ideas for future work are suggested in section VI.

## II. KEY CONCEPTS

### A. XML-based communication

#### 1) XML, DTD, and XSD

Extensible Markup Language (XML) is a data representation technique used to represent structural and hierarchical data. An XML document is composed of a set of nested nodes with only one starting node. Each node may have a number of attributes. XML document is defined by a Document Type Definition (DTD), or alternatively, an XML Schema Definition (XSD). DTDs and XSDs define the structure of the corresponding XML document, the number and type of children nodes included within any node, and some validations and constraints regarding each attribute values or possible combination of children nodes [1].

XML message structure is very lengthy and redundant. Figure 1 shows a sample Intrusion Detection Message Exchange Format (IDMEF) Heartbeat message in XML. The Bold nodes and attributes represent redundant and descriptive structure elements that are sent with each message.

#### 2) XML Schema Definition (XSD) Components

The building block in XML schema is **Element**, because it is directly mapped to an XML node. Element has a *name* attribute (representing the XML node name) and a *type* attribute (representing the XML node data type). XML schema types can be primitive types, found in XML Schema namespace, (e.g. integer, string, etc.) or new types, defined in other user-defined schemas. Schema types are categorized into two different categories; **simple types** (types composed of a single element), and **complex types** (types composed of multiple elements). Schema **Attribute** node defines an attribute of an XML node. Similar to **Element**, **Attribute** node

```xml
<?xml version="1.0" encoding="UTF-8"?>
  <idmef:IDMEF-Message version="1.0"
xmlns:idmef="http://iana.org/idmef">
    <idmef:Heartbeat messageid="abc123456789">
      <idmef:Analyzer analyzerid="hq-dmz-analyzer01">
        <idmef:Node category="dns">
          <idmef:location>Headquarters DMZ
Network</idmef:location>
          <idmef:name>analyzer01.example.com</idmef:name>
        </idmef:Node>
      </idmef:Analyzer>
      <idmef:CreateTime
ntpstamp="0xbc722ebe.0x00000000">2000-03-09T14:07:58Z
      </idmef:CreateTime>
      <idmef:AdditionalData type="real" meaning="%memused">
        <idmef:real>62.5</idmef:real></idmef:AdditionalData>
      <idmef:AdditionalData type="real" meaning="%diskused">
        <idmef:real>87.1</idmef:real></idmef:AdditionalData>
    </idmef:Heartbeat>
  </idmef:IDMEF-Message>
```

Fig. 1. Sample Intrusion Detection Message Exchange Format (IDMEF) XML – Heartbeat message (with redundant data in bold) [2].

has *name* and *type* attributes. Schema **Enumeration** node defines a single possible value for the specified type. All of the above components form the XSD, which is defined by a *Target Namespace*. It is easier to think of the *Target Namespace* as a name governing the current schema such that there should not be two similar sibling schema items of the same name. XML Schemas can reference other schemas via two types of tags / nodes; **Import**, and **Include**. Both of them has *schemaLocation* attribute, describing the location of the Schema to reference. The difference between schema **Import** and schema **Include** is schema **Import** allows importing other schemas of different target namespace, while schema **Include** allows importing schemas of the same target namespace. Therefore, schema **Import** must specify the imported schema target namespace via *namespace* attribute [1].

*B. Serialization and Deserialization*

Serialization is the process of converting complex data objects into a serial format, before sending it via transmission medium. Deserialization is the process of converting the received serial format to its original complex data objects, in order to make it ready for direct member access via code. Serial format may include Binary Stream (Byte Array), XML, or JavaScript Object Notation (JSON) [3]. In order to realize the serialization and deserialization processes, a mapping between the data object and the serial format is essential. Members that can be serialized and deserialized are marked. Serializers and Deserializers are implemented to convert data objects to and from the serial format, respectively. Examples of Serializers and Deserializers are Memory serializers (serial format is Byte Array), XML serializers (serial format is XML message), and JSON serializers (serial format is JSON message). Serialization of complex objects is done recursively for each object member, until primitive data type is found (e.g. integer, float, double, character, etc.).

*C. Cloud Computing and Service Models*

Cloud computing is based on providing consumers with different services in an elastic and measurable way. So that, consumers only pay for their usage of different computing resources. They still get the benefits of elastic resources, which can expand or shrink based on requests load. Cloud computing offers different service models. It includes Infrastructure as a Service (IaaS), Platform as a Service (PaaS), Software as a Service (SaaS), Security as a Service (SecaaS), and other Emerging Services [4].

IaaS model provides consumers with different types of resources (e.g. storage, network, and processing power). Consumers are required to build their own platform (operating system installation and configuration, and development runtime environment (RTE) installation), and application software. PaaS model is built on top of IaaS model. It provides consumers with different types of resources, and platform. Consumers are required to build their own application software. SaaS model is built on top of PaaS model. It provides consumers with different types of resources, platform, and specific software. Consumers are required to create accounts and use the offered software. Pricing is measured per account or resources usage. SecaaS model provides consumers with security-related solutions for any environment [5]; e.g. Logging Solutions (which are used to centralize logging), and Security Information and Event Management (SIEM) systems (which are complete solutions for providing security and events information storage, normalization, correlation and analysis, incident reporting, and incident interaction) [6]. Emerging service models are new services. They include Financial Software as a Service (FSaaS) model, Health Informatics as a Service (HIaaS) model, and Education as a Service (EaaS) model.

As in Figure 2, SIEM systems collect security and events information from different sources via sensors. Most information is represented in the form of formats/protocols; e.g. Syslog, IDMEF, Common Event Expression (CEE), and Simple Network Management Protocol (SNMP). Most of these protocols are based on XML [7].

Syslog is used to send log information. It is based on simple plain text; no structured format is used. It is difficult to represent structured, complex data using Syslog [8]. CEE is XML-based format, used to represent log and audit data. It also allows an organization to demonstrate compliance with audit requirements [9]. SNMP is a protocol for managing devices on IP networks. It is used for status monitoring, and configuration of network devices [10]. IDMEF is used to report an Intrusion Detection System (IDS) alert, or a device status as a heartbeat. IDMEF is based on XML. It supports structured and complex data. It also supports XML/XSD extensions, to cover any needed extra information that is not supported by the current specification of IDMEF [2]. Because of the previously mentioned benefits of IDMEF, IDMEF is selected for the study.

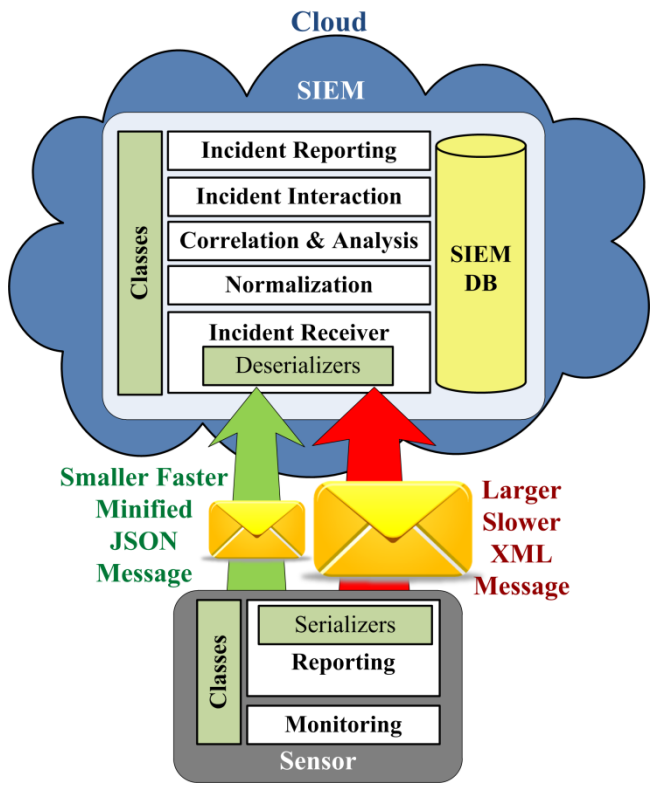

Fig. 2. Security Information & Event Management (SIEM) System Components and Communication (proposed components are in light green).

and parse by machines. It is commonly used in web systems communications. It is recommended for data communication due to its performance and message size [11] [12].

Figure 3-a shows the sample IDMEF message (of Figure 1) after conversion to JSON format. The XML message size, in Figure 1, is 686 bytes. While the JSON message size, in Figure 3-a, is 403 bytes.

JSON message is composed of a single parent object. Objects are enclosed by curly braces "{}". Objects are composed of members. Each member has a member name and value, separated by a colon ":". Member names are strings, enclosed by double quotes """. Member Values can be of simple data type, like integers (e.g. 2), strings (e.g. "dns"), or date-time (e.g. "2000-03-09T14:07:58Z"). Member Values can also be of complex data type (e.g. instance of another complex data type). Different members within an object are comma-separated ",". Array items are enclosed by square brackets "[]", with a comma separating each two consecutive items.

## III. RELATED WORK

Related work covers different topics. Attempts to reduce the unnecessary white spaces in XML are discussed. A lighter format (JSON) is used in different web systems communications. Then, the concept behind reduction in JSON is introduced. After that, the advantages and disadvantages of parsing different message types are discussed. Finally, time-consuming compression techniques are presented.

### A. XML Minification

XML messages are built based on hierarchical structure. It is common to represent them with tabs or spaces to add indentation to enhance readability. Unfortunately, these whitespace characters increase message size, regardless of the huge amount of data maintained to store structure (e.g. opening and closing tags with descriptive names).

XML Minification techniques aim to reduce message size; however, most techniques are focused on whitespace characters, and comments removal. Advanced minifiers can collapse tags that does not have content; e.g. "<idmef:real></idmef:real>" is changed to "<idmef:real/>". Examples of XML Minifiers include THE XML MINIFIER (http://www.nathanael.dk/tools_thexmlminifier.php) and WEB <MARKUP> MIN - XML Minifier (http://webmarkupmin.apphb.com/ minifiers/xml-minifier).

### B. XML vs. JSON

JavaScript Object Notation (JSON) is another data exchange format. It is lighter than XML, and easier to generate

```
{
  "IDMEF-Message": {
    "Heartbeat": {
      "messageid": "abc123456789",
      "Analyzer": {
        "analyzerid": "hq-dmz-analyzer01",
        "Node": {
          "category": "dns",
          "location": "Headquarters DMZ Network",
          "name": "analyzer01.example.com"
        }
      },
      "CreateTime": {
        "ntpstamp": "0xbc722ebe.0x00000000",
        "value": "2000-03-09T14:07:58Z"
      },
      "AdditionalData": [
        {
          "type": "real",
          "meaning": "%memused",
          "real": "62.5"
        },
        {
          "type": "real",
          "meaning": "%diskused",
          "real": "87.1"
        }
      ]
    }
  }
}
          (a)
```

```
{
  "a": {
    "a": {
      "a": "abc123456789",
      "b": {
        "a": "hq-dmz-analyzer01",
        "b": {
          "a": "dns",
          "b": "Headquarters DMZ Network",
          "c": "analyzer01.example.com"
        }
      },
      "b": {
        "a": "0xbc722ebe.0x00000000",
        "b": "2000-03-09T14:07:58Z"
      },
      "c": [
        {
          "a": "real",
          "b": "%memused",
          "c": "62.5"
        },
        {
          "a": "real",
          "b": "%diskused",
          "c": "87.1"
        }
      ]
    }
  }
}
          (b)
```

Fig. 3. Representation of IDMEF message in Figure 1: (a) JSON representation; (b) Proposed Minification with JSON representation.

*C. JavaScript Minification*

JavaScript Minification is most common in websites development and websites optimization for mobile access. It is preferred as a finalization step after development completion and before website deployment. Minification offers the following benefits: (1) File size reduction, which will minimize transmission time and network latency. (2) Faster handling and processing. (3) Minified files are better candidates to compression techniques, resulting in higher compression ratios [13]. Trivial minification includes comments, and whitespace characters removal (tabs, spaces, new lines, carriage returns, etc.). Some advanced minifiers do a more complex step, which is renaming variables, as shown in Figure 4. Examples for JavaScript Minifiers include JSCompress (http://jscompress.com), YUI Compressor (http://refresh-sf.com/yui/), and javascript-minifier (http://javascript-minifier.com).

*D. Code Generation*

Parsing and generating XML document manually is error prone. Some parsers work based on strings; e.g. element extraction is based on its name string, and setting element value is passed as a string, no matter what element data type is (http://search.cpan.org/~erwan/XML-IDMEF-0.11/IDMEF.pm)

Code generation is used to generate object oriented classes that map the corresponding XML messages based on their schemas. Messages are based on objects serialization, whereas objects are based on messages deserialization. The benefits of using code generation are: (1) Faster development time; intelligent Integrated Development Environments (IDEs) offer code auto-completion (in Microsoft Visual Studio, it is called IntelliSense), that helps developers to find the wanted member (in this case, XML element) with minimum effort. (2) Correct reference of an XML element, since elements are object's members and no strings are used. Strings are vulnerable to spelling mistakes. (3) Correct typed values assignment restricts setting each element to its value according to its element data type, rather than setting elements values as strings.

To send and receive XML messages using classes, code generation tools exist. These tools are based on the corresponding XSD. Tools for Microsoft .NET Framework include Microsoft's XSD tool (http://msdn.microsoft.com/en-us/library/x6c1kb0s%28v=vs.110%29.aspx), the open source XSD2Code (http://xsd2code.codeplex.com), etc. Tools for C++ include XSD: XML Data Binding for C++ (http://www.codesynthesis.com/products/xsd/). Tools for Java include JAXB and XmlBeans (http://www.jetbrains.com/idea/webhelp/generating-java-code-from-xml-schema.html).

*E. Compression Techniques*

XMill is a specialized XML compression technique. It compresses XML data by separating it into three components: The element and attribute names, the text values, and the tree structure of the XML document. The text values are grouped by parent element name. The three components are then compressed using standard text compression techniques [14].

Dong Zhou implemented a Structure Extraction and Encoding technique. An XML Structure is extracted; an MD5 hashing function is used to get a unique structure ID, then receiver stores Structures with their IDs in a cache. Data are sent with no structure information, associated with Structure ID only [15]. The advantage of this technique is that it works generally on any XML. The disadvantages are: (1) Similar structures are treated as new structures with new Structure ID and stored as different instances in the cache; e.g. number of items in a list, optional node or attribute, etc. (2) it is based on a cache to be available. (3) The process is considered an overhead, especially if a cache-miss occurs. A good comparison between different XML compression techniques is introduced in references [16] [17] [18].

GZip is a general-purpose compression technique. It is widely used in HTTP communication due to its good performance and high compression ratio [19] [20]. It uses the DEFLATE algorithm [21].

Compression techniques are considered a conversion process, which means it has an overhead processing time before sending the message, and after receiving the message. Direct communication techniques with message size reduction are preferred.

IV. PROPOSED SOLUTION

Proposed solution applies the JavaScript Minification techniques to XSDs, which are used to generate code that does the serialization and deserialization of objects in the minified XML format. Furthermore, it can be used with JSON serializer/deserializer, in order to make use of JSON advantages. In this case, the output will be minified JSON messages. The solution applies names minification to the underlying data format. It does not affect the generated members' names. This reduces message size but maintains software code readability. Proposed solution is implemented using Microsoft .NET Framework in C# language.

*A. Solution Architecture*

In order to achieve this goal, the solution architecture (Figure 5) shows two main tools: XSDMinify and Code Generators. (1) XSDMinify: it works on an XML Schema file and applies structure names minification. It produces two files; the first is the minified XML Schema, and the second is a dictionary file mapping each minified element name to its original element path in the original XSD (Figure 6). This process is performed only once per original XSD or any of its referenced schemas change. (2) Code Generators: XSD2Code is an open-source code generator from an XSD. It generates serializable C# classes from an XSD. Some changes are made to support the minified XSD and dictionary files as input. This tool generates serializable data fields with the minified names and getter/setter properties/methods to get/set the data fields. The properties/methods names are based on the original, meaningful and descriptive names from the dictionary file (See Figure 7). For other programming languages, the

```
function product(num1,num2)        function product(n,r){return n*r}
{ return num1*num2; }
           (a)                                   (b)
```

Fig. 4. JavaScript Minification: (a) original sample function. (b) the same function after minification.

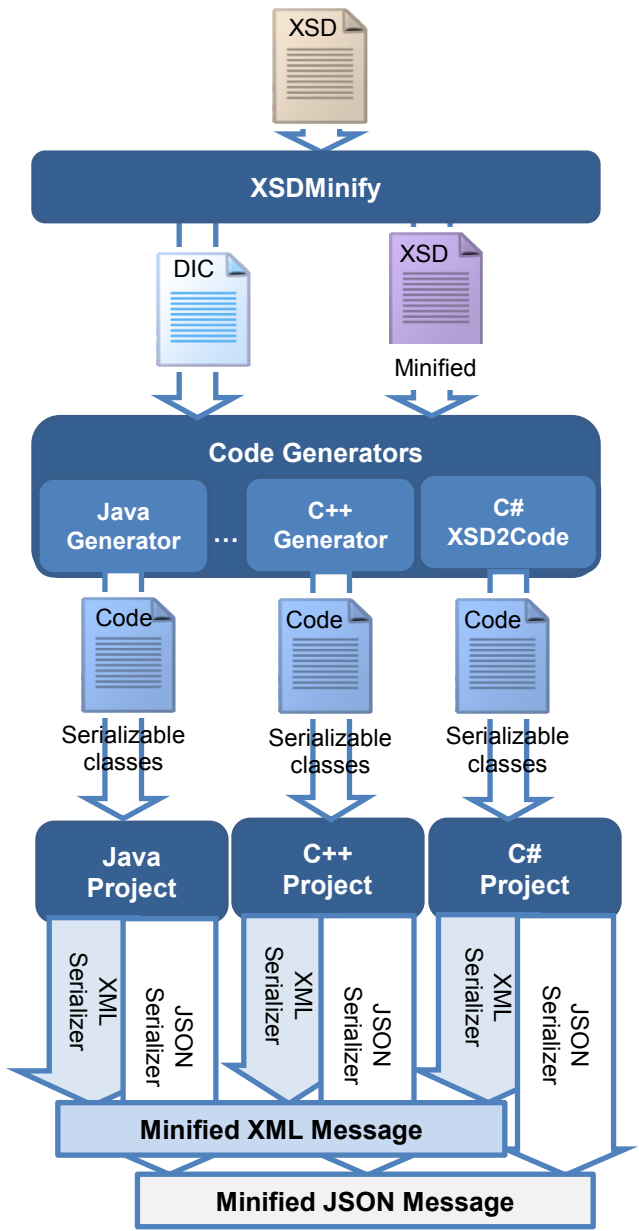

Fig. 5. Proposed solution architecture.

corresponding code generation tool needs to be customized to generate object oriented classes using the same technique.

As in Figure 2, the generated code is then included in the sender and receiver development projects. In this case, sender project represents a sensor, and receiver project represents SIEM system module. Typed messages are composed at the sender, serialized with any serializer (preferably JSON serializer), and transmitted to the receiver. The receiver

a,xsd:schema/xsd:element[name=IDMEF-Message]
b,xsd:schema/xsd:element[name=Alert]
a,xsd:schema/xsd:complexType[name=IDMEF-Message]/xsd:attribute[name=version]

Fig. 6. Sample of the dictionary (Generated from XSDMinify).

receives the message, deserializes it using the appropriate deserializer. Now the message is ready for use as an object, at receiver's side.

*B. XSDMinify*

XSDMinify is the tool that reduces the XML documents structure by applying schema structure names renaming. The original XSD file is the only input the tool requires. XSDMinify automatically detects schema Imports or schema Includes, fetches these referenced schemas, and applies minification to the referenced schemas first.

XSDMinify has two main passes for processing and minifying any XSD file. The first pass checks for schema Import or schema Include tags, then pushes the referenced schema in a stack. Therefore, the children/referenced schemas are at the top of the stack; while their parent/referencing schemas are at the bottom of the stack. The second pass represents the main minification process. In this pass, schemas are popped from the stack for minification. Schema's Target Namespace is detected, and a new Target Namespace is specified for the minified schema. Then, processing Import and Include tags is done through updating referenced Target Namespaces and schemas' new locations. This is followed by a search for any mention of the referenced schema, and an update with the minified names. After that, minification process of the current schema starts. Search for any node with "name" attribute or enumeration node with "value" attribute is carried out. An order generator generates new shorter names (e.g. a, b, c, etc. or 0, 1, 2, etc.) for each of the found name or value, respectively. Node path is also considered during order generation to allow reuse of short names. Such that, two nodes with different paths can have the same short name. A dictionary is built to store the short name mapping with the node path, and saved to a file with DIC extension (See Figure 5). Figure 6 shows a sample of the dictionary file; where short names are associated with its corresponding node path (starting from the root "schema" node to the leaf node). As processing original XSD continues, original names are replaced by the short names. Any references to the original names are updated as well. The changes are saved as the minified XSD (See Figure 5).

*C. Code generators (XSD2Code)*

XSD2Code is changed to handle code generation differently. Code Generation is based on the minified XSD and the dictionary (DIC) files, which are generated from the XSDMinify tool. As in Figure 7, Properties are generated with serializable short name fields, while property names with the

```
[DataMember(EmitDefaultValue      public enum usercategory : uint {
= false)]                              [XmlEnumAttribute("0")]
private Analyzer a;                    unknown = 0u,
public Analyzer Analyzer {             [XmlEnumAttribute("1")]
    get { return this.a; }             application = 1u,
    set { this.a = value; } }          [XmlEnumAttribute("2")]
                                       osdevice = 2u,}
        (a)                                    (b)
```

Fig. 7. Sample of the generated code using modified XSD2Code: (a) Code Generation of a Property (XSD Element); (b) Code Generation of an Enum (XSD Enumeration).

original meaningful names are accessible through code. Similarly, Enumerations are based on integer series. These integer values are used in serialization while Enumeration members are the original meaningful name. This way, a typed and meaningful access to the object's properties is achieved, resulting in maintaining software code readability. However, shorter structure elements names are used for transmission.

## V. EXPERIMENT AND RESULTS

### A. Experiment

Generated Code using proposed tool (XSD2Code), and original technique (using Microsoft's XSD tool) is included in two projects: (1) First project is a desktop application, simulating software alert source/sensor. It generates alert messages and sends them to the receiver end (the second project). (2) Second project is a web application project, simulating SIEM system, which receives alerts via a web service, processes alerts, and calculates results statistics.

Two experiments are established to compare the proposed technique's message size reduction and performance. The first experiment compares the proposed technique against traditional XML messages. The second experiment "Compression" compares the proposed technique against XMill compressed messages and GZip compressed messages.

### B. Test Data

Several types of IDMEF messages are used:

*1) Empty Message:* Almost empty message with necessary parts sent only (AnalyzerTime, CreateTime, DetectTime, and messageid fields only set.)

*2) Full Message:* IDMEF Message with all fields filled with appropriate data.

*3) Sample IDMEF Message:* IDMEF messages as represented in Examples section of the IDMEF protocol at IETF [2]. Samples are Tear Drop, ping of death, Port Scanning – 1 (Connection to a Disallowed Service), Port Scanning – 2 (Simple Port Scanning), loadmodule – 1, loadmodule – 2, phf, File Modification, System Policy Violation, Correlated Alerts, Analyzer Assessments, and Heartbeat messages.

Analysis of message structure is performed, including Raw XML Message size, Total Nodes Count for the whole message, Total Attributes Count for the whole message, and XML Complexity / Levels (representing the number of levels for nesting nodes). Table I and Figure 8 show the results of this analysis. For larger numbers, it is expected to have longer message processing time, and larger reduced message size as well.

### C. Test Environment

The sender project is hosted on a Desktop PC (Intel Pentium 4, with 3.4 GHz Processor and 3 GBs of RAM, with Network Connection of 512 Kbps Download Speed and 128 Kbps Upload Speed).

Receiver (the web services) project is hosted on Microsoft Azure Cloud Small instances. Small instance is a virtual machine with a single core 2.10 GHz processor, 1.75 GBs of memory. Instances run Microsoft Windows Server 2008 R2

TABLE I. MESSAGE STRUCTURE ANALYSIS.

| Message Type | XML Message Size (Bytes) | Total Nodes Count | Total Attributes Count | XML Complexity / Levels |
|---|---|---|---|---|
| Empty Alert | 558 | 5 | 7 | 3 |
| Complete Alert | 5219 | 107 | 70 | 6 |
| Tear Drop | 1461 | 23 | 20 | 6 |
| Ping Of Death | 1387 | 25 | 22 | 6 |
| Port Scanning 1 | 1623 | 30 | 26 | 6 |
| Port Scanning 2 | 1304 | 22 | 19 | 6 |
| Load Module 1 | 1076 | 19 | 17 | 6 |
| Load Module 2 | 1581 | 35 | 22 | 6 |
| phf | 1450 | 27 | 19 | 6 |
| File Modification | 2352 | 51 | 31 | 7 |
| System Policy Violation | 1618 | 30 | 23 | 6 |
| Correlated Alerts | 1674 | 31 | 21 | 6 |
| Analyzer Assessments | 1772 | 37 | 20 | 6 |
| Heartbeat | 736 | 11 | 9 | 5 |

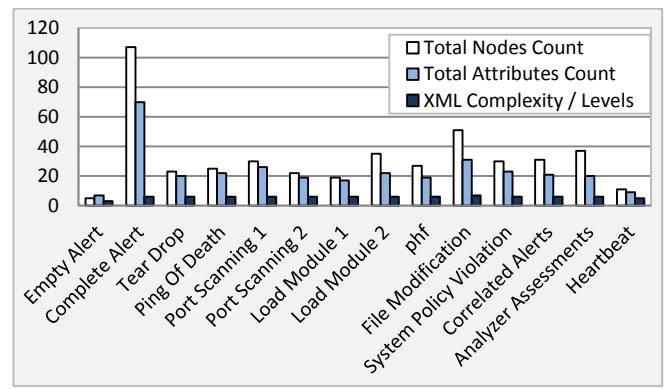

Fig. 8. Message Structure Analysis.

Enterprise Edition – 64 bit. Cloud instances use AutoScale feature for elasticity, with one to four instances. The services are hosted as Cloud Services, somewhere in West Europe.

### D. Test Results

For Experiment 1, the sender sends a burst of 500 messages. This results in total of 1000 messages for each type of the 14-message types. For experiment 2, a burst of 100 messages is sent for each message type, for each compression technique. Averages are recorded. Experiment 1 Test results for message size and transmission time (including serialization and deserialization time) (Table II) are recorded for normal XML message (Figure 1) against the proposed minified JSON message (Figure 3-b). Results show message size reduction ranging from 8.15% to 50.34%. Performance is enhanced by 35 to 342 milliseconds. The cloud instances' overall CPU usage did not exceed 5.55% of the CPU speed.

Experiment 2 results include Execution Time and Message Size analysis for compression techniques. Operations are abbreviated. Table III illustrates the abbreviations used and the function of each abbreviated process. Table IV shows Average Execution Time results. Table V shows Message Size results. Figure 9 shows combined average Execution Time for different techniques, including encoding and decoding times.

TABLE II. EXPERIMENT 1 RESULTS (SHOWING MESSAGE SIZE RESULTS IN BYTES, AND TRANSMISSION TIME RESULTS IN MILLISECONDS).

| Message Type | Message Size Results Size (Bytes) | | | Transmission Time Results | | | | | |
|---|---|---|---|---|---|---|---|---|---|
| | | | | Minimum (ms) | | Maximum (ms) | | Mean (ms) | |
| | *XML* | *Minified JSON* | *Reduction %* | *XML* | *Minified JSON* | *XML* | *Minified JSON* | *XML* | *Minified JSON* |
| Empty Alert | 558 | 450 | **19.35** | 1027 | 1010 | 8803 | 2471 | 1455 | 1420 |
| Complete Alert | 5219 | 2592 | **50.34** | 1503 | 1169 | 3782 | 2732 | 1926 | 1585 |
| Tear Drop | 1461 | 1096 | **24.98** | 1120 | 1056 | 2234 | 2166 | 1523 | 1463 |
| Ping Of Death | 1387 | 1274 | **8.15** | 1119 | 1070 | 3759 | 3825 | 1525 | 1480 |
| Port Scanning 1 | 1623 | 1061 | **34.62** | 1135 | 1046 | 5487 | 2203 | 1553 | 1460 |
| Port Scanning 2 | 1304 | 957 | **26.61** | 1110 | 1044 | 4779 | 2157 | 1520 | 1454 |
| Load Module 1 | 1076 | 894 | **16.91** | 1074 | 1046 | 2200 | 2238 | 1492 | 1449 |
| Load Module 2 | 1581 | 1092 | **30.92** | 1142 | 1061 | 2276 | 2195 | 1545 | 1463 |
| phf | 1450 | 996 | **31.31** | 1125 | 1042 | 2260 | 2178 | 1525 | 1449 |
| File Modification | 2352 | 1450 | **38.35** | 1218 | 1082 | 2343 | 2187 | 1621 | 1489 |
| System Policy Violation | 1618 | 1066 | **34.11** | 1139 | 1053 | 2270 | 2310 | 1548 | 1460 |
| Correlated Alerts | 1674 | 1185 | **29.21** | 1142 | 1061 | 2282 | 2216 | 1550 | 1470 |
| Analyzer Assessments | 1772 | 1195 | **32.56** | 1160 | 1061 | 2259 | 2252 | 1558 | 1468 |
| Heartbeat | 736 | 404 | **45.10** | 1047 | 1005 | 2385 | 3647 | 1453 | 1418 |

TABLE III. EXPERIMENT 2 "COMPRESSION" ABBREVIATIONS.

| Process Abbreviation | Description |
|---|---|
| XML | Serialization of traditional XML messages. No compression used. |
| De XML | Deserialization of traditional XML messages (inverse of the "XML" process). |
| XMill | Compression of XML messages using the specialized XMill compressor. |
| De XMill | Decompression of XML messages using the specialized XMill compressor (inverse of the "XMill" process). |
| GZip XML | Compression of XML messages using GZip compressor (a cyclic redundancy check value for detecting data corruption is included). |
| De GZip XML | Decompression of XML messages using GZip compressor (inverse of the "GZipXML" process). |
| Min JSON | (Proposed Technique) Serialization into Minified JSON messages. No compression used. |
| De Min JSON | (Proposed Technique) Deserialization from Minified JSON messages (inverse of the "MinJSON" process). |
| GZip Min JSON | (Proposed Technique) Serialization into Minified JSON messages, plus using GZip compression (a cyclic redundancy check value for detecting data corruption is included). |
| De GZip Min JSON | (Proposed Technique) Decompression of the compressed Minified JSON messages (inverse of the "GZipMinJSON" process). |

TABLE IV. EXPERIMENT 2 AVERAGE EXECUTION TIME IN MILLISECONDS.

| Message Type | De XML | XML | GZip XML | GZip Min JSON | De GZip XML | Min JSON | De Min JSON | De GZip Min JSON | De XMill | XMill |
|---|---|---|---|---|---|---|---|---|---|---|
| Empty Alert | 0.01 | 0.04 | 0.15 | 0.01 | 0.04 | 0.08 | 0 | 1.12 | 10.6 | 17.3 |
| Complete Alert | 0.27 | 0.08 | 1.27 | 1.02 | 1.17 | 4.81 | 2.42 | 3.51 | 16.4 | 25.4 |
| Tear Drop | 0.09 | 0.07 | 0.21 | 0.07 | 0.24 | 0.59 | 0.61 | 1.02 | 11.9 | 18.9 |
| Ping Of Death | 0.05 | 0.13 | 0.17 | 0.13 | 0.19 | 0.16 | 1.09 | 1.25 | 11.7 | 18.8 |
| Port Scanning 1 | 0.04 | 0.49 | 0.05 | 0.17 | 0.14 | 0.07 | 0.97 | 1.06 | 11.6 | 19.3 |
| Port Scanning 2 | 0.17 | 0.02 | 0.13 | 0.18 | 0.5 | 0.02 | 0.37 | 1.56 | 11.8 | 18.7 |
| Load Module 1 | 0.03 | 0.07 | 0.23 | 0.16 | 0.19 | 0.05 | 0.28 | 1.08 | 12 | 19.5 |
| Load Module 2 | 0.23 | 0.01 | 0.28 | 0.2 | 0.22 | 0.02 | 1.3 | 1.17 | 11.3 | 19.1 |
| phf | 0.02 | 0.34 | 0.1 | 0.08 | 0.19 | 0.04 | 0.6 | 1.07 | 11.8 | 18.7 |
| File Modification | 0.07 | 0.05 | 0.17 | 0.39 | 0.96 | 0.07 | 1.15 | 1.97 | 11.8 | 20.2 |
| System Policy Violation | 0.01 | 0.16 | 0.16 | 0.37 | 0.28 | 0.21 | 1.19 | 1.1 | 12.6 | 19.1 |
| Correlated Alerts | 0.11 | 0.06 | 0.04 | 0.17 | 0.21 | 0.4 | 1.21 | 1.65 | 11.1 | 19.1 |
| Analyzer Assessments | 0.22 | 0.03 | 0.46 | 0.11 | 0.6 | 0 | 1.21 | 1.35 | 11.8 | 19.7 |
| Heartbeat | 0.18 | 0.25 | 0.11 | 0.64 | 0.01 | 0.01 | 0.03 | 0.11 | 11.2 | 17.9 |

TABLE V. EXPERIMENT 2 MESSAGE SIZE IN BYTES.

| Message Type | GZipMinJSON | GZipXML | XMill | MinJSON | XML |
|---|---|---|---|---|---|
| Empty Alert | 345 | 420 | 436 | 452 | 574 |
| Complete Alert | 888 | 1387 | 1659 | 2594 | 5488 |
| Tear Drop | 620 | 784 | 822 | 1098 | 1532 |
| Ping Of Death | 633 | 772 | 774 | 1276 | 1471 |
| Port Scanning 1 | 633 | 832 | 834 | 1063 | 1719 |
| Port Scanning 2 | 601 | 757 | 757 | 959 | 1379 |
| Load Module 1 | 566 | 714 | 724 | 896 | 1136 |
| Load Module 2 | 626 | 800 | 802 | 1094 | 1681 |
| phf | 601 | 787 | 794 | 998 | 1536 |
| File Modification | 677 | 953 | 980 | 1452 | 2501 |
| System Policy Violation | 614 | 835 | 855 | 1068 | 1713 |
| Correlated Alerts | 650 | 815 | 814 | 1187 | 1766 |
| Analyzer Assessments | 658 | 949 | 958 | 1209 | 1888 |
| Heartbeat | 415 | 559 | 556 | 406 | 772 |

XMill takes the longest execution time. However, other techniques take much shorter execution time between 0.6 and 1.6 milliseconds. Figure 10 shows detailed average Execution Time for different techniques. The prefix "De" signifies Decoding/Decompression Times, while the un-prefixed techniques signify Encoding/Compression Times.

Figure 11 shows Average Message Size for different techniques. Compared to XML, GZipped Minified JSON Messages are 66.1% shorter. GZipping the original XML files produces 54.8% shorter messages. The time-consuming specialized XMill compressor produces 53.22% shorter messages. Raw minified JSON messages are 37.37% shorter, without applying any compression. Figure 12 shows the detailed message size comparison for all experiment techniques.

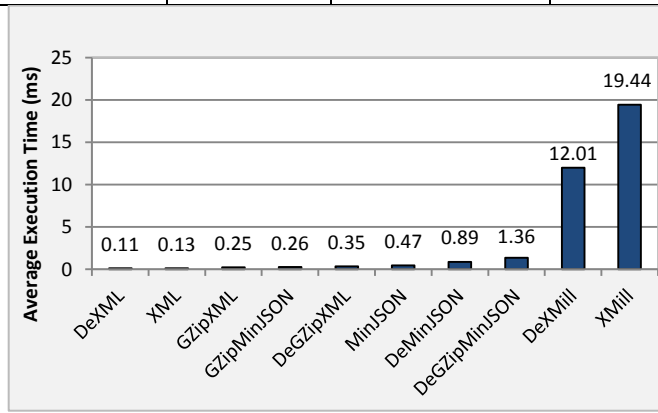

Fig. 10. Experiment 2: Results Comparison of Average Execution Time for Different Techniques in milliseconds.

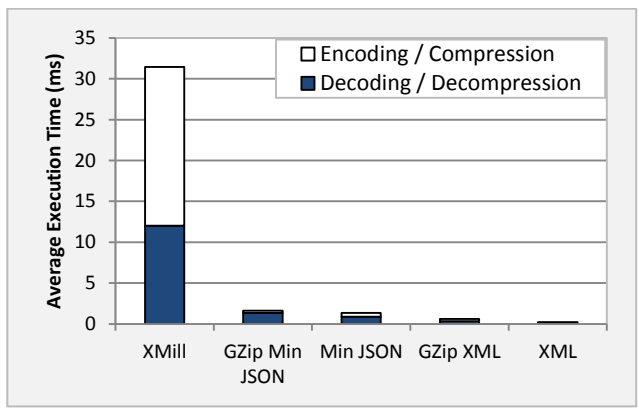

Fig. 9. Experiment 2: Results Comparison of Average Execution Time in milliseconds of Messages Encoding and Decoding using different techniques.

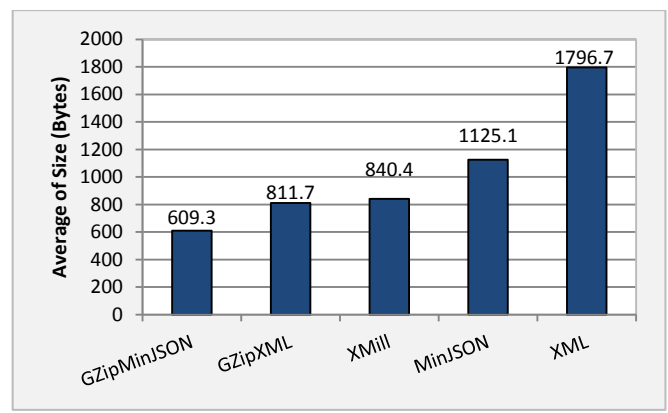

Fig. 11. Experiment 2: Results Comparison of Average Message Size for Different Techniques in Bytes.

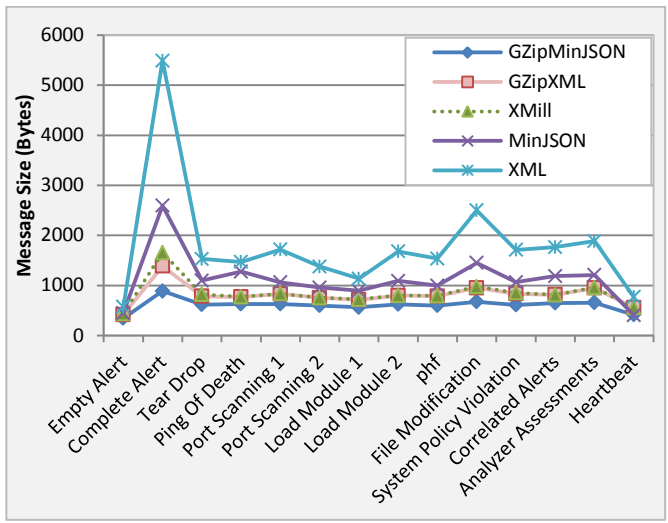

Fig. 12. Experiment 2: Results Comparison of Message Size in Bytes for different message types, for all techniques.

## VI. Discussion and Conclusion

In this paper, we introduced a new XML Schema-based minification and communication technique in JSON message format. XML Schemas are minified using XSDMinify tool. This process is required only once per XML schema change. Then, the generated minified XML schema is processed with customized XSD code generation tool (XSD2Code). The code generation step generates code that sends and receives shorter minified messages. Based on the serialization type, communication can occur using shorter XML messages, or even shorter JSON messages. We performed our tests on Microsoft Azure Cloud platform, using different IDMEF messages types. Experiment 1 results show message size reduction ranging from 8.15% to 50.34% compared to raw XML messages. Performance is enhanced by 35 to 342 milliseconds. This technique is applied to raw messages, without applying any compression techniques (like those techniques introduced in section III). Compression techniques yield better results because of the similarities found in the new message structure (e.g. the minified names alphabets (a, b, c, …, and 1, 2, 3, …) instead of the full meaningful names). Experiment 2 applies both XML Compression Technique, and the general purpose GZip compression technique. As average results for all message types, XMill compression produces 53.2% shorter message, but XMill is very expensive in Execution Time (takes 31.45 extra milliseconds). Applying the proposed Minified JSON technique yields 37.37% shorter message compared to original XML messages. Minified JSON technique has extremely low execution time, reaching 1.36 milliseconds only. Adding GZip compression to Minified JSON technique produces 66.1% shorter message size compared to original XML messages (with 12.9% shorter size compared to XMill). GZipping Minified JSON technique takes 1.62 milliseconds only (94.85% faster than XMill).

To conclude, the proposed technique "Minified JSON messages" is a better alternative to using traditional XML messages, or specialized XMill compression. This technique produces a reasonable message size reduction, with almost no performance overhead. To achieve the best results, incorporation of GZip compression and Minified JSON technique is recommended. This produces the ultimate compression ratio, with a tiny negligible performance overhead. A separation between the names of the object oriented classes' members and the underlying transmission representation is well-established to maintain code readability. For future work, well-defined procedure for incorporating XML extensions will be studied. Data visualization tools may be considered for adopting the generated dictionary file (resulting from the minification process), in order to visualize minified data for user viewing.